\documentclass{desyprocA4}
\usepackage{amssymb}
\usepackage{graphics}
\usepackage{epsfig}

\def\beq{\begin{equation}}
\def\eeqn{\end{equation}}

\newenvironment{Eqnarray}%
   {\arraycolsep 0.14em\begin{eqnarray}}{\end{eqnarray}}
\def\beqa{\begin{Eqnarray}}
\def\eeqa#1{\label{#1}\end{Eqnarray}}
\def\eeqan{\end{Eqnarray}}

\def\leqsim{\mathbin{\;\raise1pt\hbox{$<$}\kern-8pt\lower3pt\hbox{$\sim$}\;}}
\def\geqsim{\mathbin{\;\raise1pt\hbox{$>$}\kern-8pt\lower3pt\hbox{$\sim$}\;}}

\def\XPM#1{\mbox{$ \tilde{\chi}^{\pm}_#1                                $}}
\def\XN#1{\mbox{$ \tilde{\chi}^0_#1                                     $}}

\def\p#1{\mbox{$ \mbox{\bf p}_1                                         $}}

\newcommand{\smur}    {\mbox{$ \tilde{\mu}_{\mathrm R}                     $}}

\newcommand{\sel}     {\mbox{$ \tilde{\mathrm e}                           $}}
\newcommand{\sell}    {\mbox{$ \tilde{\mathrm e}_{\mathrm L}               $}}
\newcommand{\selr}    {\mbox{$ \tilde{\mathrm e}_{\mathrm R}               $}}

\newcommand{\snu}     {\mbox{$ \tilde\nu                                   $}}

\newcommand{\sq}     {\mbox{$ \tilde{q}                                $}}

\newcommand{\stau}    {\mbox{$ \tilde{\tau}                                $}}
\newcommand{\stone}   {\mbox{$ \tilde{\tau}_1                              $}}

\newcommand{\stq}     {\mbox{$ \tilde {\mathrm t}                          $}}
\newcommand{\stqone}  {\mbox{$ \tilde {\mathrm t}_1                        $}}

\newcommand{\sbq}     {\mbox{$ \tilde {\mathrm b}                          $}}

\newcommand{\eeto}    {\mbox{$ {\, \mathrm e}^+ {\mathrm e}^- \to             $}}

\newcommand{\ba}{\begin{array}}
\newcommand{\ea}{\end{array}}
\newcommand{\bc}{\begin{center}}
\newcommand{\ec}{\end{center}}
\newcommand{\be}{\begin{eqnarray}}
\newcommand{\eeq}{\end{eqnarray}}
\newcommand{\bes}{\begin{eqnarray*}}
\newcommand{\ees}{\end{eqnarray*}}
\newcommand{\Kz}{\ifmmode {\rm K^0_s} \else ${\rm K^0_s} $ \fi}
\newcommand{\Zz}{\ifmmode {\rm Z^0} \else ${\rm Z^0 } $ \fi}
\newcommand{\xxbar}{\ifmmode {\rm x\bar{x}} \else ${\rm x\bar{x}} $ \fi}
\newcommand{\rphi}{\ifmmode {\rm R\phi} \else ${\rm R\phi} $ \fi}

\def    \missEt      {\ifmmode{/\mkern-11mu E_t}\else{${/\mkern-11mu E_t}$}\fi}
\def    \missE       {\ifmmode{/\mkern-11mu E}\else{${/\mkern-11mu E}$}\fi}
\def    \missp       {\ifmmode{/\mkern-11mu p}\else{${/\mkern-11mu p}$}\fi}
\def    \misspt      {\ifmmode{/\mkern-11mu p_t}\else{${/\mkern-11mu p_t}$}\fi}

\begin{document}
\title{Simplified SUSY at the ILC}

\author{{\slshape Mikael Berggren$^1$}\\[1ex]
$^1$DESY, Notkestra{\ss}e 85, 22607 Hamburg, Germany\\
}

\desyproc{DESY 13-136}
\doi  

\maketitle


\begin{abstract}
At the ILC, one has the possibility to search for SUSY in an model-independent way:
The corner-stone of SUSY is that sparticles couple as particles. This is independent of
the mechanism responsible for SUSY breaking. Any model will have one Lightest SUSY
Particle (LSP), and one Next to Lightest SUSY Particle (NLSP). In models with conserved
R-parity, the NLSP must decay solely to the LSP and the SM partner of the NLSP. Therefore, 
studying NLSP production and decay can be regarded as a ``simplified model without
simplification'': Any SUSY model will have such a process.

The NLSP could be any sparticle: a slepton, an electroweak-ino, or even a squark.
However, since there are only a finite number of sparticles, one can systematically search 
for
signals of all possible NLSP:s. This way, the entire space of models that have a kinematically
reachable NLSP can be covered. For any NLSP, the ``worst case'' can be determined,
since the SUSY principle allows to calculate the cross-section once the NLSP nature and
mass are given. The region in the LSP-NLSP mass-plane where the ``worst case'' could be
discovered or excluded experimentally can be found by estimating background and efficiency
at each point in the plane. From experience at LEP, it is expected that the lower signal-to
background ratio will indeed be found for models with conserved R-parity.

In this document, we show that at the ILC, such a program is possible, as it was at
LEP. No loop-holes are left, even for difficult or non-standard cases: whatever the NLSP is
it will be detectable.
\end{abstract}
%
\section{Introduction}
\label{sec:intro}

Simplified SUSY models are models where one only considers the direct decay
of a sparticle to it's standard model partner and the lightest SUSY particle, the
LSP.
Or, stated somewhat more generally, it can be defined as searching for SUSY in 
the generic topology
``SM particles + missing energy``. The missing energy does not necessarily
need to be due only to the undetected LSP, but can also have contributions
from cascade decays where the respective mass-differences in the steps of
the cascades are too small for the decays to be detected.

At LHC, simplified models have become a widely used and important tool to cover the more 
diverse phenomenology beyond 
constrained SUSY models. 
However they come with a substantial number of caveats themselves, 
and great
care needs to be taken when drawing conclusions from limits based on the simplified approach. 
LHC is particularly powerful in searching for colored
sparticles, either gluinos or first and second generation squarks.
Exactly these particles are those that are expected to be the
the heaviest sparticles in most SUSY models.
Therefore it is rather unlikely that their decays directly to
the LSP and the SM partner would have a large branching ratio,
and that the ``simplified model'' approach would un-ambiguously cover
a large parameter-space in a general MSSM model.

%


At lepton colliders, the situation is quite different.
In such machines, simplified models gives the 
possibility to search for SUSY in an model-independent way:
The corner-stone of SUSY is that {\it sparticles couples as particles} \cite{Dimopoulos:1981zb}. 
This is independent of the
mechanism responsible for SUSY breaking. In particular, the couplings to the photon and
the Z-boson are known,
which implies that the cross-section for any \eeto ~sparticle -- anti-sparticle pair process 
that proceeds only via the s-channel is determined by the kinematics alone, 
ie. by $\sqrt{s}$ and the mass of the sparticle.
Contributions from sparticle exchange in the t-channel are possible
in $\snu_e , \sel, \XN{i}$ or $\XPM{i}$ pair-production,
by exchange of  a $\XPM{i} , \XN{i} , \sel $ or $\snu_e$, respectively.
Also with such diagrams contributing, all couplings are known,
and the cross-section will still be determined by the kinematic relations,
at the cost that one must consider both the mass of the produced and of the exchanged sparticle.

Furthermore, by definition there is one LSP, and one NLSP. If stable, it is un-avoidable that
the LSP must neutral and weakly interacting, due to cosmological constraints. The NLSP, on
the other hand, could be any sparticle - a slepton, a bosino, or even a squark. However, there is
only a limited number of sparticles. While an arbitrary sparticle in the spectrum of any specific
SUSY model typically would decay through cascades of other, lighter sparticles, the NLSP only
has one decay-mode, namely to the LSP and the SM partner of the NLSP,
ie. exactly the topology defining a simplified model. 

In non-minimal models with more neutralinos 
(eg. the nMSSM \cite{Fayet:1974pd}),
it could be that the NLSP couples only very little to most particles and sparticles.
In the nMSSM, this would be the case if the NLSP is mostly singlino. 
The first directly detectable sparticle would be instead be the nest-to-next to lightest sparticle (the NNLSP). 
But, since the
NLSP hardly couples to the NNLSP, the NNLSP would decay directly to the
LSP and it's SM partner with branching ratio close to 100 \%,
ie. the experimental signature is unchanged. 
If instead the LSP couples very little, eg. if the LSP is the singlino of nMSSM,
or if it is the gravitino, as is common in gauge-mediated SUSY breaking (GMSB)\cite{Giudice:1998bp},
the picture does not change compared to the MSSM: the NLSP still can only decay
to the LSP and it's SM-partner. The smallness of the coupling might even help, as
it could imply that the NLSP will decay at a detectable distance from the production
point, a very clean experimental signature.

Therefore, studying
NLSP production and decay can be regarded as a ``simplified model without simplification'': 
Any SUSY model will have such a process.
Putting these observations together, one realizes that by systematically searching for signals
for all possible NLSP:s, at all possible values for $M_{NLSP}$ and $M_{LSP}$,
the entire space of models that are reachable can be covered. At the ILC,
such a program is possible {\it without leaving loop-holes}, even for difficult or non-standard cases.

In fact, a program of this kind was already carried out in the past, at LEP.
Even though 
the LEP combinations compiled by the LEPSUSYWG \cite{LEPSUSYWG}, as well
as the publications of the individual 
experiments \cite{Heister:2002hp,Abdallah:2003xe,Achard:2003ge,Abbiendi:2003sc}
put some
emphasis on interpreting the negative results in terms of the then-popular
mSUGRA \cite{Chamseddine:1982jx} or CMSSM \cite{Kane:1993td} models, they do provide
exactly such a set exclusion regions in the $M_{NLSP}$--$M_{LSP}$ plane for all possible NLSP:s.
Among the individual LEP experiments,
most notably DELPHI published  all R-parity conserving searches in a single 
paper \cite{Abdallah:2003xe}.

In the following, we will first discuss the generic procedure to follow to cover all
possible, kinematically accessible cases,
and how to evaluate what LSP-NLSP masses would be
possible to discover (if realized in nature), or to excluded 
(if not). 
We will then comment on a number of potentially difficult cases that might present
loop-holes, and finally present a realistic simulation study of the
expected discovery and exclusion reaches for $\smur$ and $\stone$ NLSP:s
at the ILC.

\section{Covering all possibilities}

\begin{figure}[htb]
  \begin{center}
  \includegraphics[width=0.425\textwidth]{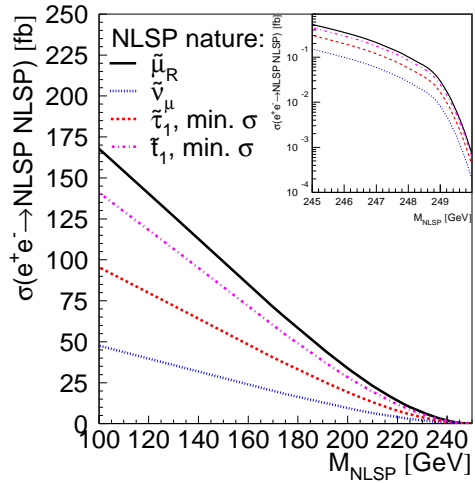}
  \end{center}
  \caption{NLSP production cross-sections as a function
of M$_\mathrm{NLSP}$ at  $\sqrt{s}$ = 500 GeV
for a few NLSP
hypotheses. The beam-spectrum was the taken from the ILC TDR specifications,
and the beam polarization was +80\% (-30\%) for the electron (positron) beam.
The insert shows the last few GeV before the
kinematic limit, on a log-scale.
}
\label{fig:xsect}
\end{figure}

A full scan over  the $M_{NLSP}$--$M_{LSP}$ plane for any possible NLSP is 
done in a number of steps.
First, the cross-section at a chosen $\sqrt{s}$ for 
NLSP pair-production is determined from
the SUSY-principle and kinematics. 
The cross-section only depends on which sparticle the
NLSP is and it's mass (and in some cases of the properties of a
sparticle contributing to a t-channel diagram, as mentioned above); 
it does not depend on the properties of
the LSP.
Figure \ref{fig:xsect} shows the cross-section 
at  $\sqrt{s}$ = 500 GeV 
as a function of  $M_{NLSP}$ for a selection of NLSP-candidates.
These cross-sections were computed using {\tt whizard} v. 1.95 \cite{Kilian:2007gr}.

The best analysis for each of the (potentially) different
signal regions is then determined. 
To do so, generated signal events in a few typical points for the given NLSP (high,
intermediate and low $M_{NLSP}-M_{LSP}$ ($=\Delta(M)$), close to or far from threshold) 
are passed through full detector
simulation. Also simulation of the background from all SM-processes at each mass-point is
done at this stage.
The analysis exploits the typical features of NLSP production and decay at
an $e^+e^-$-collider, namely: missing energy and momentum, high acolinearity, correct particle/jet
identification, invariant di-jet/di-lepton mass-conditions, possibly using 
constraint kinematic fitting. 
A very powerful feature due to the known initial state at the ILC is that the kinematic
edges of the detected systems can be precisely calculated at any 
point in the $M_{NLSP}$--$M_{LSP}$
plane. In particular, close to kinematic limit where the width of the decay-product spectrum
is quite small, this feature allows for an almost background-free signal with high efficiency. 
At
this stage, the background at each point can be estimated by passing the correctly normalized
samples of simulated events from all SM processes through the analysis chain.

Finally, signal samples are needed over a fine grid in 
the $M_{NLSP}$--$M_{LSP}$ plane, in order to evaluate the
signal efficiency. 
Full detector simulation is much too time-consuming 
to allow for a fine enough
grid of points. Instead, fast simulation is tuned to  reproduce the fully simulated
signal in the typical points as well 
as possible, 
and used to extrapolate to the full plane. 
Note that for background, only  the analysis of the events
differs from point to point.
Hence, there is no
need to generate or simulate SM-events separately at each point; 
one only needs to apply the
analysis relevant for the point. 
Therefore, it is foreseeable to evaluate the SM background using
full detector simulation.
The number of signal and background events at each 
point are used to calculate the 
significance of the signal at the given integrated luminosity. 
It can be interpreted either as discovery potential ie. 
how much bigger than the background the signal 
would be in terms of number of standard deviations of the 
background fluctuation, or as exclusion - 
ie. how much smaller than 
the
signal added to the background the background 
alone would be in terms of number of 
standard deviations of the background+signal fluctuation. 
In points where the former is bigger than 5$\sigma$, one would expect to 
discover the signal at the 5$\sigma$ level, 
if it is indeed there, 
while if the latter
is bigger than 2$\sigma$, one would
expect to be able to exclude the signal at 95 \% CL, 
if there is only background.
In the former case, the point is within {\it Discovery Reach}, 
in the latter within {\it Exclusion Reach}. 
As this is procedure is executed for {\it every possible NLSP}, 
it will constitute a completely model-independent search for SUSY.

\section{A few difficult cases: Do they present loop-holes ?}

The discussion above might give rise to some objections, since it contains a number of assumptions
that are true in many models, but not all. For instance, R-parity might be broken, yielding an
LSP that decays. 
If the LSP is a higgsino, a singlino or a gravitino, the NLSP-LSP coupling might be so small that the 
NLSP decays after a detectable distance in the
the detector, or even outside the detector.
Sparticles might be mixed states of different SM-partners. The mass-difference
between the LSP and the NLSP might be very small (making the decay of the NLSP hard to
observe). 
The decay of the NLSP might be invisible.
There might be two or more states very close in mass, making the uniqueness of the
NLSP decay-signal doubtful. In the following we will address these issues.

\subsection*{Unusual decays: RPV, gravitino LSP, ...}

In R-parity violating SUSY, the LSP need not be stable nor neutral.
If the LSP is long-lived and charged, or if it has a life-time such that it often decays within the
detector, the search-reach would be even better than in the R-conserving case. 
The same would be the case if instead the NLSP is long-lived and charged, as might occur
if the LSP is the gravitino, or is a singlino, or if the NLSP--LSP mass difference is lower than the mass of
the SM partner of the NLSP.
If the LSP is
long-lived, but neutral, the experimental signature is the same as in the R-conserving case.
In RPV SUSY,
the only case of concern is therefore the one where the LSP-decay is prompt. This will make
the experimental situation more complex. Various combinations of $\lambda$, $\lambda '$, and
$\lambda ''$ are strongly
constrained by other observations from quark and lepton flavor physics. Therefore, many different
cases, with different signatures needs to be studied. Nevertheless, experience from LEP indicates
that also in this case, the searches will be more sensitive 
than in the 
R-conserving case \cite{Heister:2002jc,Abdallah:2003xc,Achard:2001ek,Abbiendi:2003rn}.

\subsection*{Mixed NLSP:s}
\subsubsection*{Mixed sfermions}

In the case of a sfermion NLSP which is a mass-state mixed between the L and R hyper-charge
states, one will have one more parameter - the mixing angle\footnote{Note that for the sneutrinos, only the
L state exists, so mixing only occurs for charged sfermions.}. However, as the couplings to the Z
of both states are known from the SUSY-principle, so is the coupling with any mixed state. There
will then be one mixing-angle the represents a possible ``worst case'', which allows to determine
the reach whatever the mixing is - namely the reach in this ``worst case''. 
This is illustrated in Figure \ref{fig:xsect}, where the shown cross-sections for $\stone$ and $\stqone$
are those corresponding to the mixing-angles yielding the lowest possible cross-section.
It should be noted
that one can't mix away $\eeto \tilde{f} \tilde{f}$ completely: The coupling to $Z$ might vanish, but not to $\gamma$.

In the case  the NLSP is the $\sel$ or the $\snu_e$, there is a t-channel contribution to the production cross-section,
via $\XN{1}$ or $\XPM{1}$ exchange, respectively. 
This contribution always interferes  constructively with the s-channel.
The Yukawa coupling of both $e$ and $\nu$ is quite small (and consequently that of $\sel$ and $\snu_e$, by
the SUSY principle). One can minimize the cross-section by ``switching off'' the t-channel contribution by assuming that 
the $\XN{1}$ or the $\XPM{1}$ are pure higgsinos, and thereby also in this
case arrive at a well defined ``worst case''.
\subsubsection*{Mixed bosinos}

A more complicated case is the mixed bosino NLSP. Here the mixing-structure is more complex,
and usually the there will be correlations between the nature of the LSP and the NLSP. Therefore,
both cross-sections and decay products are influenced by the mixing, and it will be hard to set
limits in the $M_{NLSP}$--$M_{LSP}$ plane. In this case, one would need to somewhat 
lower the ambition
and evaluate reachable cross-sections instead. These cross-sections can then be converted to
reach in the $M_{NLSP}$--$M_{LSP}$ plane for any specific model. However, 
a number of generic observations can be made:
\begin{itemize}

\item The $\XPM{1}\XPM{1}$ cross-section can only be suppressed if there is destructive 
interference between
the s-channel and the $\snu_e$ mediated t-channel. The latter only becomes 
important if $\snu_e$ is
close in mass to $\XPM{1}$, and does not contribute at all if $\XPM{1}$ 
is higgsino-like, due to small Yukawa coupling of the neutrinos.
In addition,  $\snu_e$ is the partner of $\nu_{eL}$, so one would expect $\sell$
to have about the same mass as the (light)  $\snu_e$. 
In most models, $\selr$ is expected to be even lighter than $\sell$, so in this
chargino-suppressed case, it would be likely that the NLSP is the  $\selr$, not
the $\XPM{1}$.

\item In the higgsino region (low $\mu$), $\XPM{1}$ and $\XN{2}$ are close in mass.

\item In the higgsino region, the cross-sections for $\XN{2}$ pair-production 
and $\XN{1}\XN{2}$ associated production cannot both be suppressed at the same time.

\item In the gaugino region, $\XN{2}$ production proceeds via a t-channel 
$\sel$ exchange, and will be
suppressed if the $\sel$ is very heavy. 
However, in this region it is expected that $\XPM{1}$ is close in
mass to, or lighter than, the $\XN{2}$. 
Therefore, not seeing {\it any} bosino would imply that at the same
time $\snu_e$ is light - in order to suppress $\XPM{1}\XPM{1}$ - 
and $\sel$ is heavy - to suppress $\XN{2}\XN{2}$, which is very difficult to
accommodate in any model.
\end{itemize}
Taken together, these observations indicate that within the MSSM it is very
unlikely that the NLSP is a bosino that could escape detection at the ILC:
either the NLSP would not a bosino, but rather a slepton, or the production 
cross-section of the NLSP (or possibly a nearby NNLSP) 
would be sizable.

\subsection*{Very low $\Delta(M)$}

In the case of a very small mass differences between the LSP and the NLSP - less than a few
GeV - the clean environment at the ILC nevertheless allows for a good detection efficiency. If
$\sqrt{s}$ is much larger than the threshold for the NLSP-pair production, the NLSP:s 
themselves will
be highly boosted in the detector frame, and most of the spectrum of the decay products will
be easily detected. In this case, the background will come from pair-production of the NLSP's
SM partner. Therefore the precise knowledge of the initial state is of paramount importance to
recognize the signal, by the slight discrepancy in energy, momentum and acolinearity between
signal and background.

In the case the threshold is not much below $\sqrt{s}$ the background from 
$\gamma\gamma \rightarrow f \bar{f}$ where
the $\gamma$:s are virtual ones radiated of the beam-electrons becomes severe. 
The beam-remnant electrons themselves are
deflected so little that they leave the detector un-detected through the out-going beam-pipes.
This background can be kept under control by demanding that there is a visible ISR photon
accompanying the soft NLSP decay products. If such an ISR is present in a $\gamma\gamma$
event, 
the beam-remnant
will also be detected, and the event can be rejected \cite{bib:higgsino_paper}.
In the case of a $\tilde{q}$ NLSP, the decay of the NLSP will be complicated 
if  $\Delta(M) < m_q$, but still
governed by known physics of many-body decays. An open question is R-hadrons, ie. the case
when the low phase-space for the $\sq$-decay allows the $\sq$ to hadronize before decaying. 
How such a
particle would interact with the detector-material can only be speculated about,
but would most probably stand out as being very different from any SM particle.

\subsection*{Invisible NLSP decays}

In some cases the decay of the NLSP might be
un-detectable. 
The prime example of this is the case of a $\snu$ NLSP, where the decay would be
$\snu \to \nu \XN{1}$. 
It could also arise eg. in nMSSM or GMSB models with a very long-lived neutralino NLSP. 
In such cases, the techniques presented in \cite{Bartels:2009fa} can be used: by demanding the
presence of an ISR photon and nothing else in the event, one can scan the effective
center-of-mass energy ($\sqrt{s'}$), defined by
$$
s' = s - 2 \sqrt{s} E_{\gamma}, 
$$
where $\sqrt{s}$ is the nominal center-of-mass energy,
and $E_{\gamma}$ is the energy of the ISR photon.
The NLSP-pair production can be observed as 
an excess of events in the $d\sigma/d\sqrt{s'}$ distribution 
above the irreducible SM background from $\eeto \nu\bar{\nu}\gamma$.
The onset of the excess will allow to determine the NLSP mass, and the
shape will contain information about the partial wave the pair is
produced in, which in turn will determine whether a pair of scalars
or fermions were produced.
Furthermore, the difference in the dependence on beam-polarization between eg. 
$\snu$  and $\XN{2}$ pair-production
will give further handles in determining the properties of the NLSP.

It should also be pointed out that this technique could be used to observe
pair-production of the LSP itself.
\subsection*{Degeneracy, ie  $>$ 1 NLSP}

The case of (near) degenerate NLSP:s is not so much a problem with the detection of the signal
- more open channels will yield higher total signal - but of it's interpretation. As SUSY makes
definitive predictions of cross-sections, also a {\it too large} signal cross-section might 
lead to the conclusion that new physics 
is discovered, but it cannot be SUSY. To make a correct interpretation,
it is therefore important to be able to separate close states.
In the case several sleptons are close to degenerate in mass, and form an NLSP-group would
not pose any mayor experimental problems as the different decays are very well separable by
the detector. 
Also the case if $\sbq$ and $\stq$ would form a close to degenerate NLSP-group
can be separated experimentally, albeit not so efficiently as sleptons.

If the degeneracy is not complete, but there still is some small 
difference in mass
between the states, the interpretation might be complicated by cascade-decays of the slightly
heavier state to the lighter one+neutrinos. In this case, the capability to precisely choose the
beam-energy at the ILC will make it possible to study the ``real'' NLSP below the threshold of
it's nearby partner.

Several studies both for small and large mass-differences have shown that the most likely case
of degenerate NLSP:s, namely  $\XPM{1}$-$\XN{2}$ degeneracy can be well separated 
experimentally at the ILC \cite{bib:higgsino_paper,SueharaList}.

\section{Simulation study at ILC}

A concrete example on how to cover
all possibilities is presented below. 

For the simulation, an integrated 
luminosity of 500 fb$^{-1}$
at E$_{CMS}$ = 500 GeV was assumed. 
The average polarizations of the electron and positron beams
were assumed to be +80\% and -30\%, respectively. 
The detector simulation used was the fast
simulation program SGV \cite{SGV},
adapted to the ILD detector concept at ILC \cite{bib:ILDDBD}.
Beam-conditions as those
described in the ILC TDR were assumed \cite{bib:ILCTDR}. 
The signal was generated using {\tt pythia} v. 6.422 \cite{bib:pythia}.
The background SM samples generated for the detector benchmarking \cite{bib:ilc-whizard}
for the ILC TDR produced
using {\tt whizard} 1.95 \cite{Kilian:2007gr} were used. 
Under these conditions the total expected SM background is 1270 million events.

\subsection*{NLSP independent step}

To select events for further study, a set of pre-selection cuts aimed at 
reducing the SM background
by one to two orders of magnitude, while retaining a efficiency close to 100 \% for the signal were
applied. 
These strongly reduced background samples could be used,
without loss of generality, when studying any NLSP. 
While the cuts were not aimed at a particular NLSP hypothesis, nor a particular point in the
mass-plane, 
it is nevertheless useful to already at this stage separate low- and high-multiplicity
final states into two separate streams. 
In the following, further details of the low multiplicity case will be given.

\subsection*{Low multiplicity final states}

For the search for NLSP:s with low multiplicity of the SM decay products, 
the following cuts were used 
\begin{itemize}
\item At most 10 charged particles reconstructed,
\item Missing energy ($E_{miss}$) above 100 GeV,
\item Absolute values of the cosine of the polar angle of missing momentum 
($| \cos{ \theta_{p_{miss}}} |$) below 0.95,
\item Particles clustered in 2 or 3 jets using the DELPHI tau-finder \cite{Abdallah:2003xe},
\item Visible mass ($M_{vis}$ ) below 300 GeV.
\end{itemize}

Already at this point, 91 \% of the the SM background was removed (120 million remaining),
while the the signal efficiencies typically remained above 90 \%.

\subsection*{NLSP dependent step: The slepton example}

The following step was to refine the selection depending of the NLSP.
 
The most important criterion at this stage was that the identity of the
detected particles should be in accordance of the nature of the NLSP,
but
also polar angle distributions could be exploited, using the knowledge on the spin of 
the NLSP. 

For the case of a slepton NLSP, the fact that sleptons are scalars allows
select centrally produced, non-back-to-back events with no forward-backward asymmetry :
\begin{itemize}
\item The sum of the product of the charge and the cosine of the polar angle of the two
most energetic jets should be above -1 (this cut removes most of the highly asymmetric
$W W \rightarrow l\nu l\nu$ background),
\item Not more than 2 GeV should be detected at angles lower than 20 degrees to the beam-axis.
\item The missing transverse momentum ($p_{T miss}$) should be above 5 GeV.
\item $M_{vis}$ should be more than 4 GeV away from $M_Z$.
\item The angle between the two most energetic jets, projected on the plane 
perpendicular to the beam (the acoplanarity angle, $\theta_{acop}$ ) should be 
between 0.15 and 3.1 radians,
\end{itemize}

\begin{figure}[htb]
  \begin{center}
\includegraphics[width=0.425\textwidth]{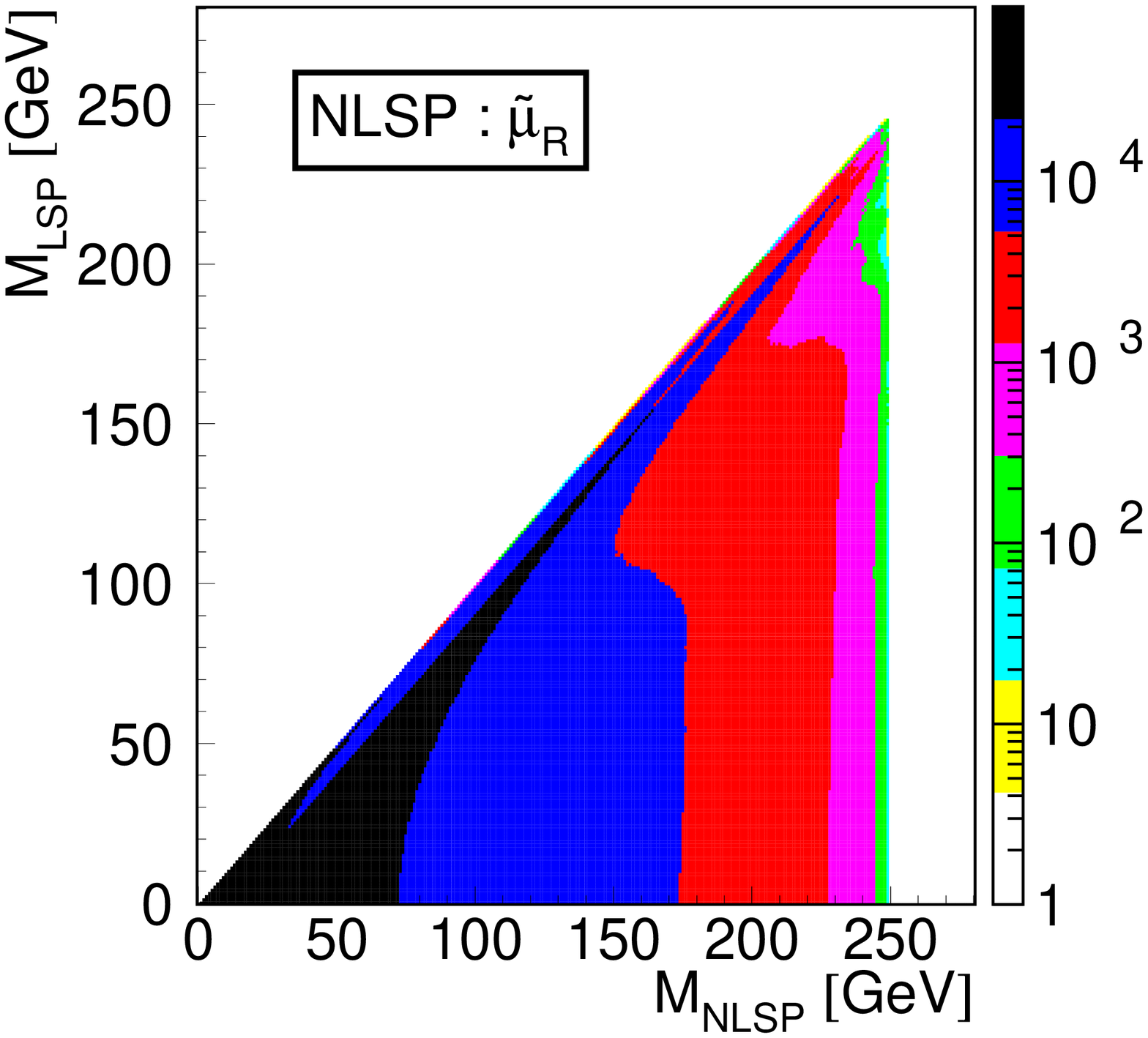}
\hspace{0.2cm}
\includegraphics[width=0.425\textwidth]{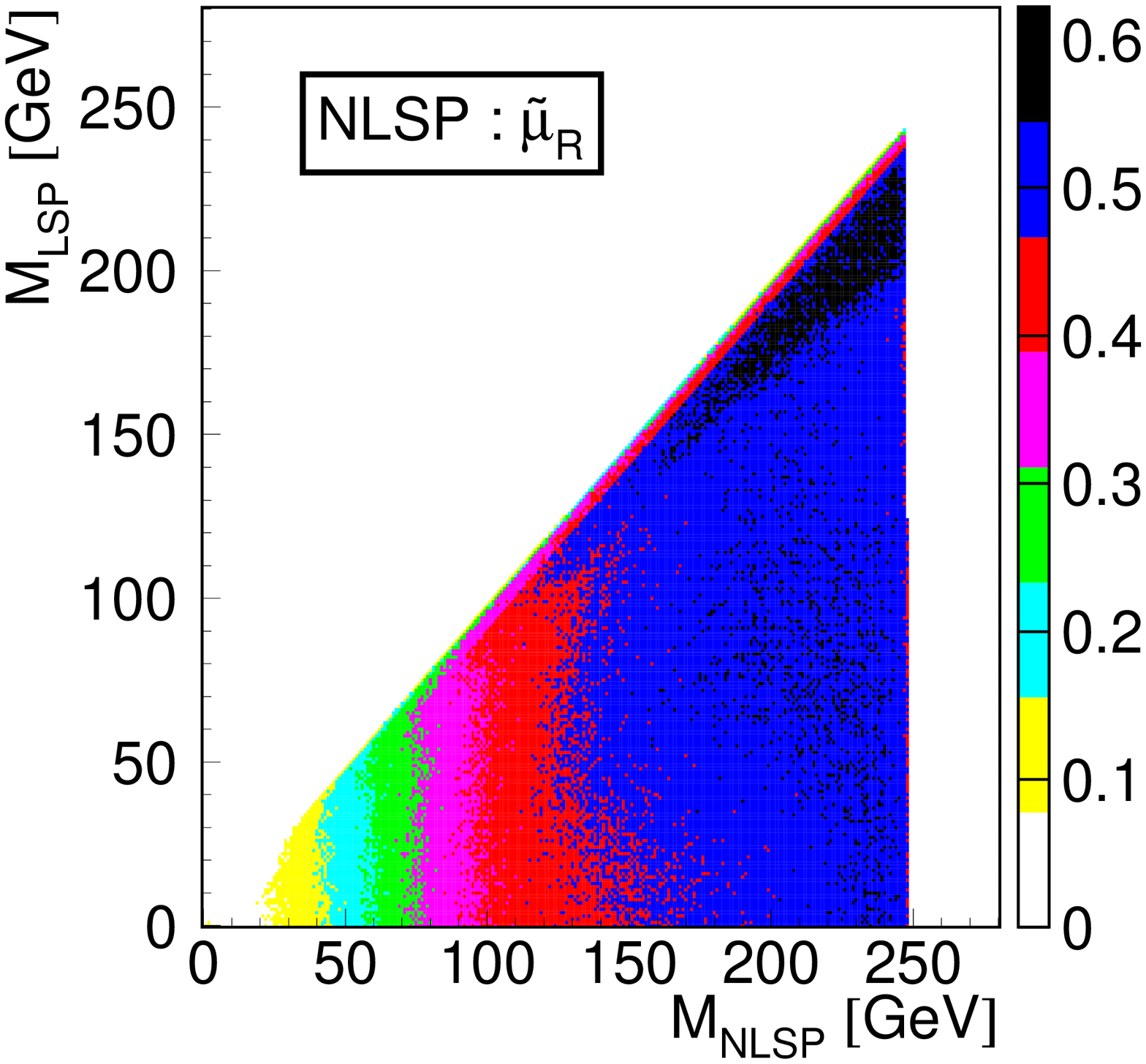}
  \end{center}
  \caption{ Left: Final background in the $\smur$ search, Right: Final signal efficiencies 
in the $\smur$ search.
}
\label{fig:smubckneff}
\end{figure}
\begin{figure}[htb]
  \begin{center}
\includegraphics[width=0.425\textwidth]{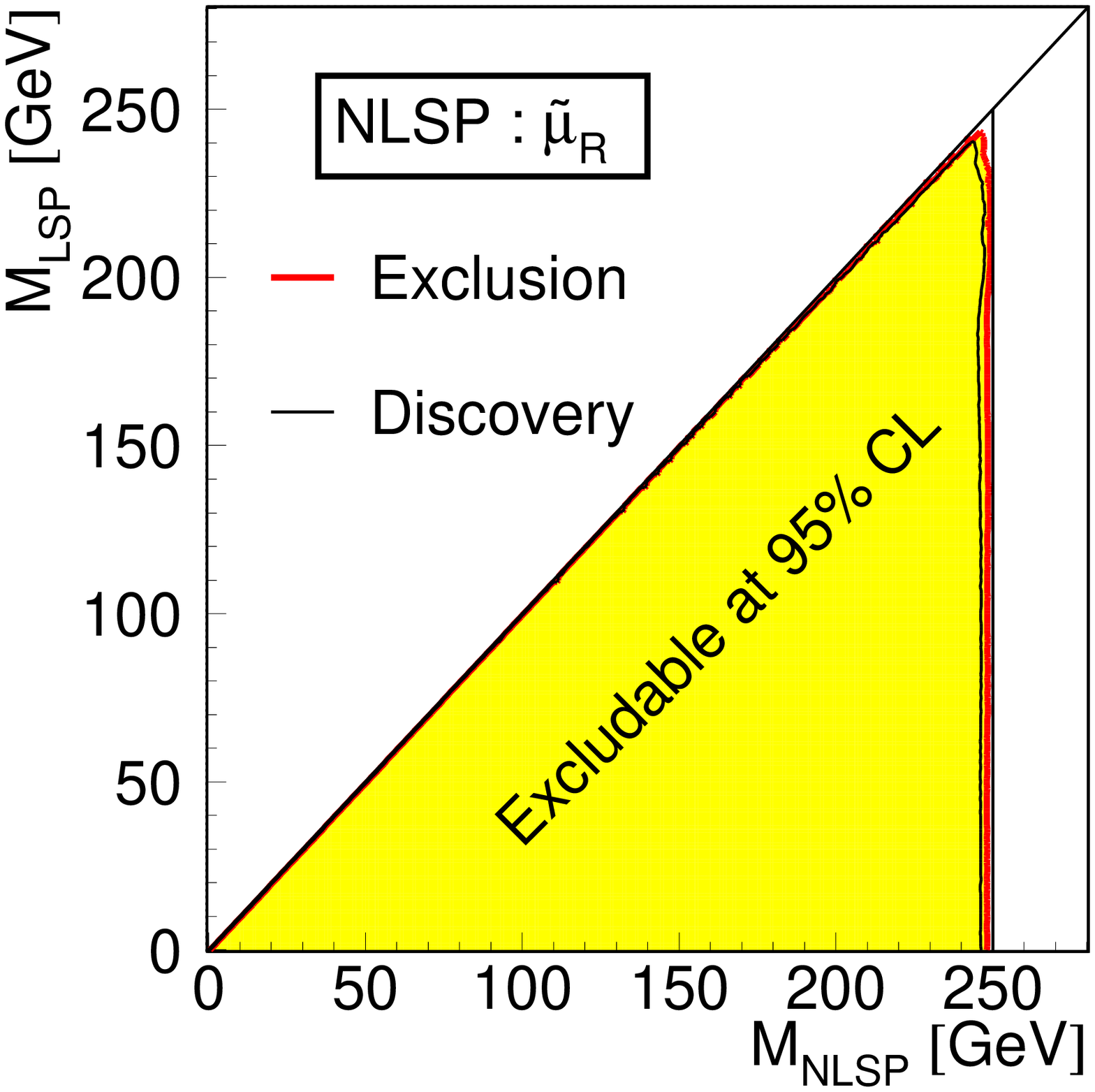}
\hspace{0.2cm}
\includegraphics[width=0.425\textwidth]{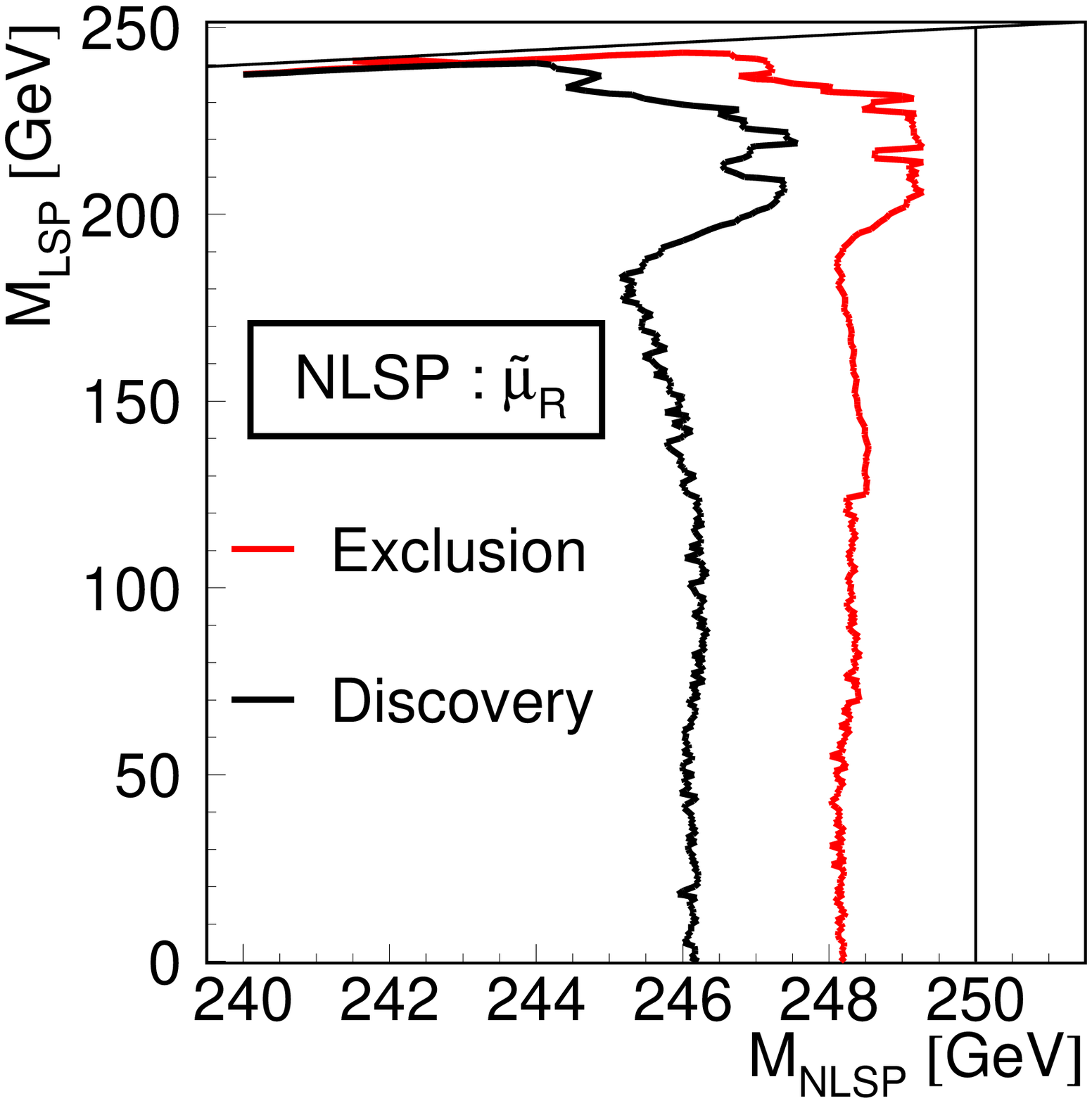}
  \end{center}
  \caption{Discovery-reach for  a $\smur$  NLSP after 
collecting 500 fb$^{-1}$
at $\sqrt{s}$ = 500 GeV. Left: full scale, Right: zoom to last few GeV before the
kinematic limit.
}
\label{fig:smuexcl}
\end{figure}

Particle identification was then  applied. In the case of a $\smur$ NLSP,
this was straight-forward: The two most energetic charged tracks should be identified as muons.
Muon identification at the ILC is expected to have an efficiency above 97 \%,
with quite low fake rate \cite{bib:ILDDBD}.
At this point, the SM background in the  $\smur$ case was about 
120 000 events (completely dominated
by $\gamma\gamma$ events), 
while the signal efficiency was still between 45 \% and 70 \% at all mass-points.

In the case of a $\stone$ NLSP, identification is much more complex, and
inevitably has a cost both in efficiency and mis-identification. 
The correct pattern of charged tracks from $\tau$-decays was first imposed:
\begin{itemize}
\item Exactly two of the jets should contain charged particles
 (allowing for one additional jet if it only contains neutral particles).
\item In each of the charged jets, there should be 1 or 3 charged particles, and
the total charge in each of them should be $\pm$1.
\item The two jets should have opposite charge.
\end{itemize}
Then a set of cuts to reduce background from sources containing
leptons that do not come from $\tau$-decays was applied:
\begin{itemize}
\item The two charged jets should not both be made of single leptons
of the same flavor.
\item To reduce the background from single W production in e$\gamma$ events 
(with W$\rightarrow \tau \nu$), and since this study was done with the 
right-handed electron, left-handed positron beam polarization,
none of the jets should be made of a single positron.
\item To further reduce this background, background from $WW \rightarrow e\nu_e \tau\nu_{\tau}$
or from $\gamma\gamma$ events with one beam-remnant deflected at large angles,
the most energetic jet should not be a single electron. 
\end{itemize}

The remaining background after these cuts was 1.7 million events,
which, as in the $\smur$ case, were completely dominated 
by $\gamma\gamma$ events.
The signal efficiency ranges from 5 \% at low $\Delta(M)$ up to 35 \% at high $\Delta(M)$.

At this stage, cuts depending on the point in the  $M_{NLSP}$--$M_{LSP}$ plane
were applied.
At each point, only events with missing mass above 2$M_{LSP}$ were accepted.
Both for $\smur$ and $\stone$, events where the
most energetic jet had an energy above the upper edge
of the decay spectrum at the given point were rejected.
For $\smur$, also events where the least energetic jet had an
energy below the lower end-point could be rejected; for $\stone$
such a cut was not possible, due to the invisible energy of the
neutrinos from the $\tau$ decay.
Only few signal events were removed by this requirement, 
while the background is strongly reduced. 
 
The signal efficiency was determined by generating 1000 events at each of the the considered mass
point (in total about 30 000 points), and passing them through the fast simulation. 
The number of background events was small enough at this stage that the important 
information for each of
them could simultaneously be stored in memory, and hence the number of background
events passing the
point-dependent cuts could be found rapidly. 
A final set of cuts was applied that depended on
the difference between the LSP and NLSP masses:
If this difference was below
10 GeV, most remaining background came from the  $\gamma\gamma$  process. 
Therefore, if $\Delta(M)$ was below 10 GeV, the following anti-$\gamma\gamma$
cuts were applied:
\begin{itemize}
\item $| \cos{ \theta_{p_{T miss}}} | < 0.7$,
\item  $\theta_{acop} < 2.8$.
\end{itemize}
In the opposite case, a cut was applied against back-to-back events:
\begin{itemize}
\item The absolute value of the cosine of the angle between the most 
energetic jets ($| \cos{ \theta_{acol}} |$)
should be below 0.85\footnote{This cut helps to reduce remaining two-fermion background. 
It is of little use in the soft region since the
background in that region is normally not back-to-back due to beam-remnants 
or ISR escaping in the beam-pipe.}.
\end{itemize}
In the case of the $\smur$ NLSP, these were the final cuts.
The left plot of Figure \ref{fig:smubckneff} shows the map of the remaining background in
this case, 
while the right one shows
the signal efficiency.
Combining the two maps with the production cross-section allows to draw the map of the expected
significance of the signal, either as exclusion-reach or as discovery-reach. 
The contours of these two regions are shown in Figure \ref{fig:smuexcl}.

In the case of the $\stone$ NLSP, further cuts are needed to reject background
events with two $\tau$:s.
Events with  two $\tau$:s back-to-back in the transversal projection can still
fake the typical signal topology of large missing transverse momentum and
high acoplanarity, namely if the two $\tau$:s decay asymmetrically, in the sense that
in one of the decays, the {\it visible} system has a direction close to the decaying $\tau$, while
in the other one the {\it invisible} neutrino is close to the 
direction of it's parent. 
Such a configuration yields one high energy jet close to the direction of one
of the $\tau$:s, and one low energy jet with little correlation to the direction
of the other $\tau$. 
Two variables in the projection perpendicular to the beam
are used to discriminate against such topologies:
$\rho$, the transverse momentum wrt. the trust-axis,
and the sum of the transverse momentum of each jet wrt. to the other jet.
In the low(high) mass difference-region, $\rho$ should be above 
3(10) GeV. In the low region, the transverse momentum sum should be below  25 GeV,
while it should be above 15 GeV in the high region.

\begin{figure}[htb]
  \begin{center}
\includegraphics[width=0.425\textwidth]{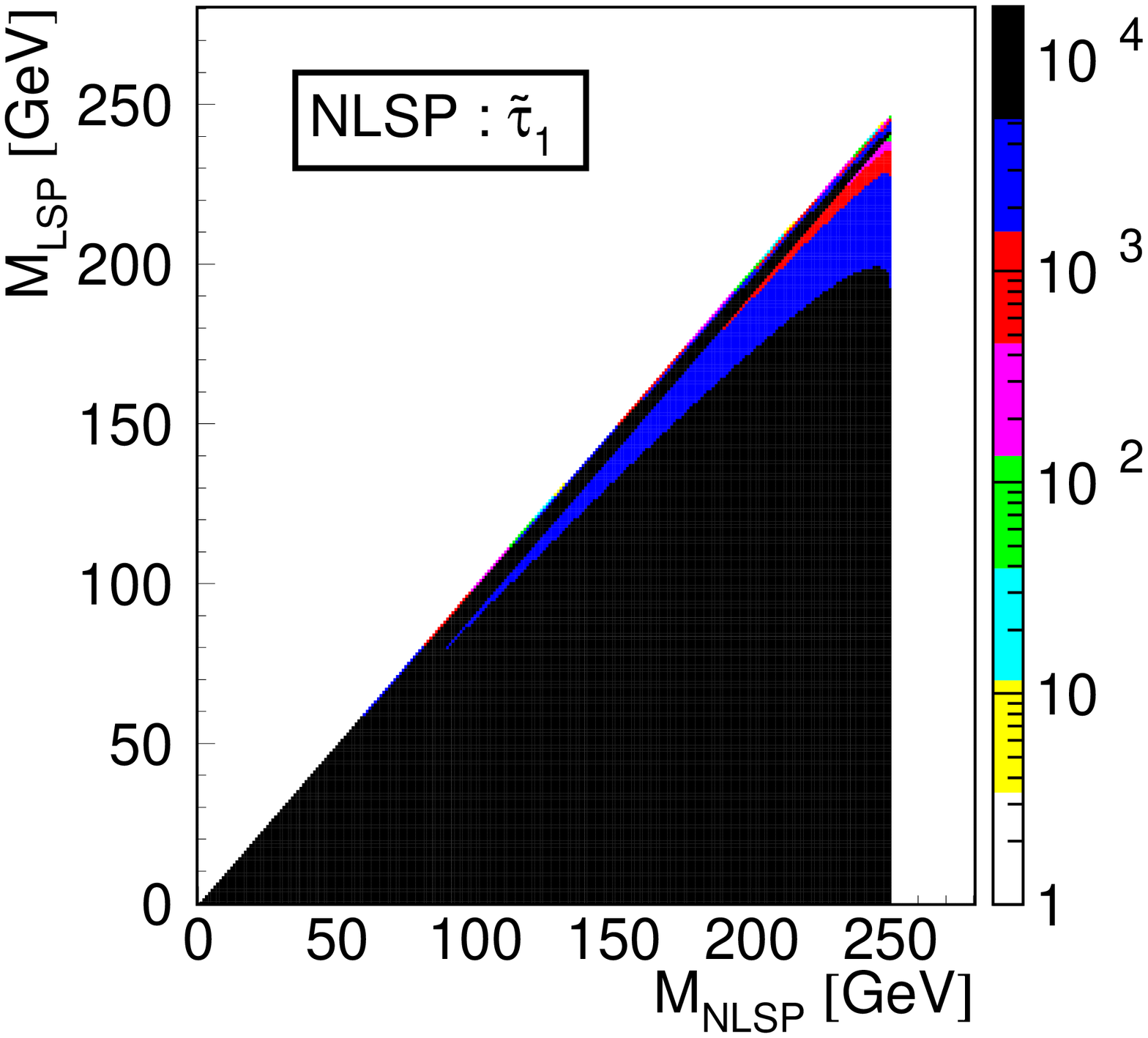}
\hspace{0.2cm}
\includegraphics[width=0.425\textwidth]{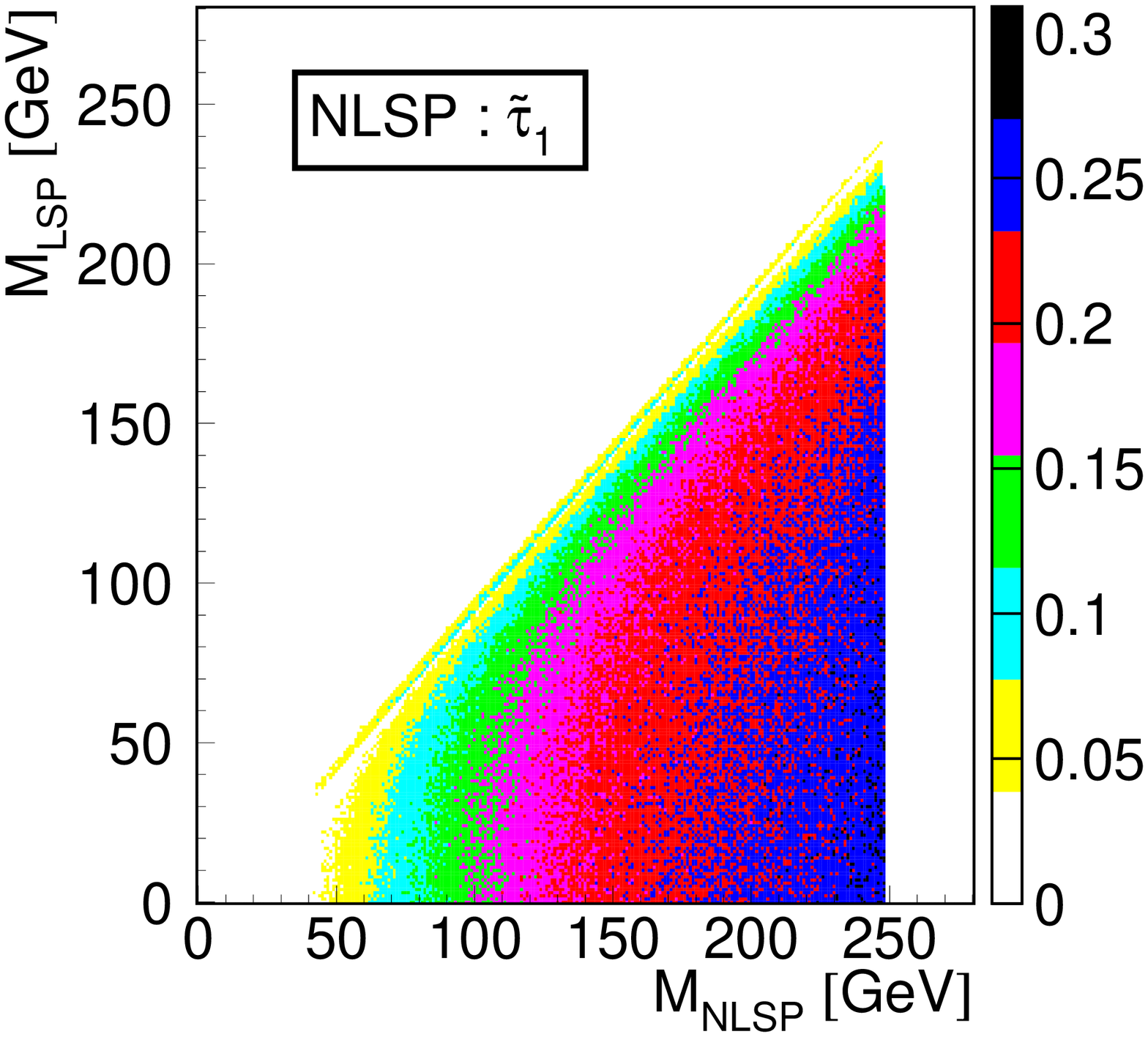}
  \end{center}
  \caption{ Left: Final background in the $\stone$ search, Right: Final signal efficiencies 
in the $\stone$ search.
}
\label{fig:staubckneff}
\end{figure}
\begin{figure}[htb]
  \begin{center}
\includegraphics[width=0.425\textwidth]{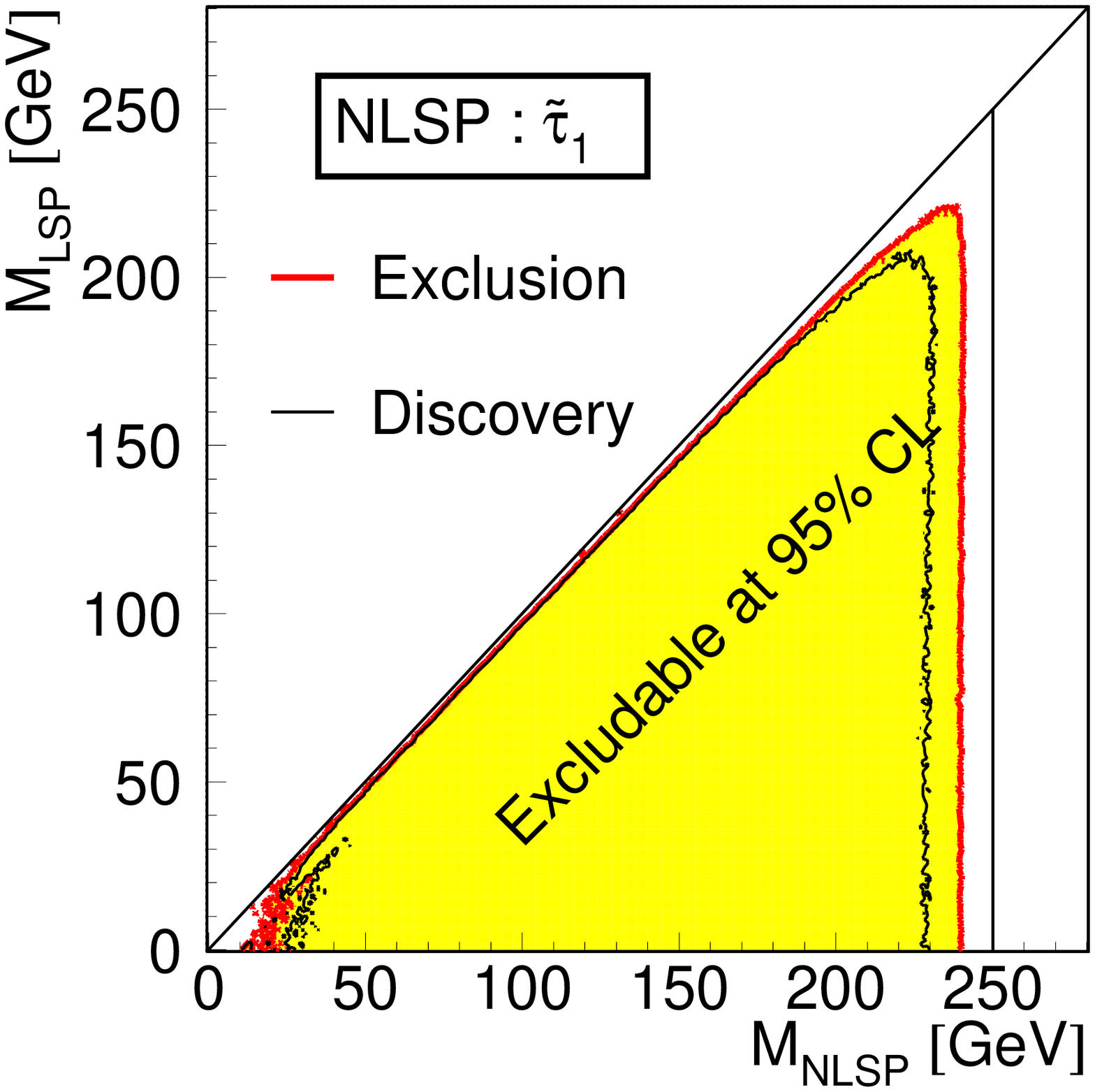}
\hspace{0.2cm}
\includegraphics[width=0.425\textwidth]{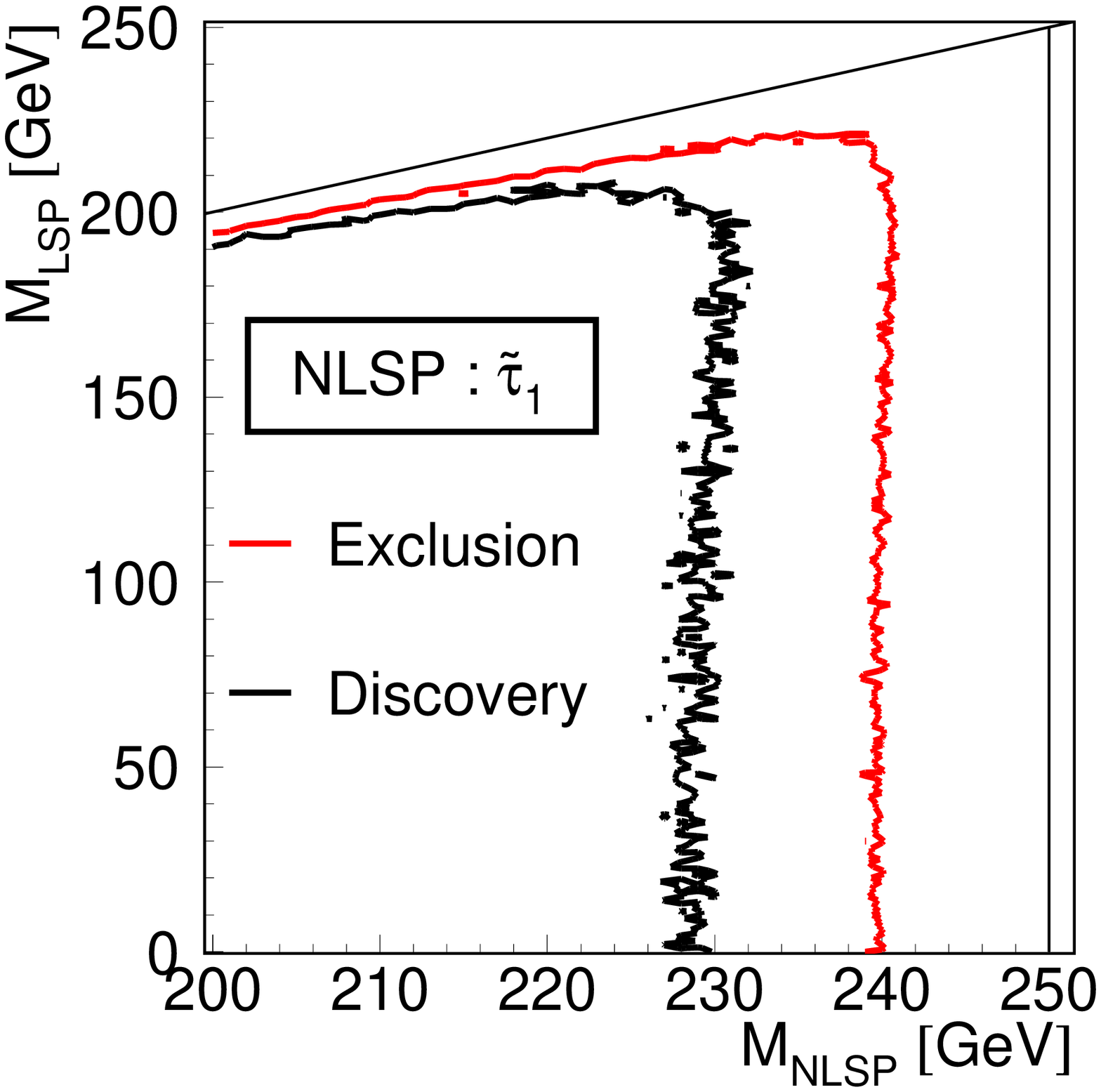}
  \end{center}
  \caption{Discovery-reach for  a $\stone$  NLSP after 
collecting 500 fb$^{-1}$
at $\sqrt{s}$ = 500 GeV. Left: full scale, Right: zoom to last few ten GeV before the
kinematic limit.
}
\label{fig:stauexcl}
\end{figure}

In Figure \ref{fig:staubckneff}, the remaining background 
and the signal efficiency is shown for the $\stone$ NLSP case.
The expected
exclusion-reach and discovery-reach 
are shown in Figure \ref{fig:stauexcl}.
As expected, these regions are somewhat smaller than in the $\smur$ NLSP case,
due to the lower production cross-section (Figure \ref{fig:xsect}),
the lower efficiency due to the much lower efficiency in identifying
$\tau$:s compared to $\mu$:s, and the higher background, due both to
the higher fake-rate in particle identification, and the fact the
point-dependent lower jet energy cut is cannot be used in the $\stone$ case.
The uncovered region at low $M_{\stone}$, which would need a special
treatment,  is not of concern since it has already been excluded by LEP \cite{Abdallah:2003xe}.

\section{Summary}
\label{sec:summary}

We have argued that, at ILC, one can  search for SUSY in an model-independent way,
by systematically searching for all possible NLSP:s over
the $M_{NLSP}$--$M_{LSP}$ plane:
Because the corner-stone of SUSY is that {\it sparticles couples as particles},
SUSY predicts the coupling to the Z and the $\gamma$ of any sparticle.
The production cross-section of any sparticle-pair produced
in the s-channel therefore only depends on kinematics, ie. the sparticle mass and  $\sqrt{s}$.
Since the NLSP can only decay to it's SM-partner and the LSP,
both cross-section and event topology is exactly predicted at each
point in $M_{NLSP}$--$M_{LSP}$ plane, independent of the mechanism
of SUSY-breaking. 

We have also discussed a number of problematic cases,
and concluded that none of them will represent a loop-hole where
SUSY can hide.

We presented the details on how the program can be carried out,
and described a realistic simulation study of two specific cases at the ILC:
Either the quite easy case where the $\smur$ is the NLSP, 
or the difficult one where the NLSP is the $\stone$, at
the $\stau$ mixing-angle yielding the smallest cross-section.
It was found that the only two different analysis procedures were
needed to cover all parts of
the  $M_{NLSP}$--$M_{LSP}$ plane, one for $\Delta(M) > 10$ GeV, one for
$\Delta(M) < 10$ GeV.
In addition, the same procedure could be used both for $\smur$ and
$\stone$, except for the different particle identification procedure,
and a set of additional cuts needed against the $\gamma\gamma$ background
for $\stone$, due to the fact that the lower kinematic edge of the
sparticle decay products cannot be
used as a discriminator when the SM partner itself decays partly to
invisible neutrinos.

The conclusion is that for the $\smur$, either exclusion or
discovery is expected up to a few GeV below the kinematic
limit (ie. $M_{NLSP}=\sqrt{s}/2$).
Also for $\stone$, one would come close to this limit: Exclusion
is expected up to 10 GeV below $\sqrt{s}/2$, while discovery would be
expected up to 20 GeV below $\sqrt{s}/2$.

\section{Acknowledgments}
We would like to thank the ILC Generators group  
for providing the SM background samples.
We also thank
Gudrid Moortgat-Pick,  Jenny List, Howard Baer and 
Werner Porod for helpful discussions.
We thankfully acknowledge the support
by the DFG through the SFB 676 ``Particles, Strings and the Early Universe''.

\section{Bibliography}
\bibliographystyle{unsrt}
\begin{footnotesize}

\end{footnotesize}

\end{document}